\documentclass[twocolumn]{aastex63}

\usepackage{lineno}
\usepackage{multirow}
\usepackage{booktabs}
\usepackage{graphicx}   
\usepackage{hyperref}   

\newcommand{\nn}{\nonumber}
\submitjournal{ApJ Letters}

\shortauthors{Roy Choudhury}

\begin{document}

\title{\boldmath Cosmology in Extended Parameter Space with DESI Data Release 2 Baryon Acoustic Oscillations: A 2$\sigma$+ Detection of Nonzero Neutrino Masses with an Update on Dynamical Dark Energy and Lensing Anomaly}
\email{
	sroy@asiaa.sinica.edu.tw
}

\author{Shouvik Roy Choudhury}

\affiliation{Institute of Astronomy and Astrophysics, Academia Sinica, No. 1, Section 4, Roosevelt Road, Taipei 106319, Taiwan}

\begin{abstract}

We obtain constraints in a 12 parameter cosmological model using the recent Dark Energy Spectroscopic Instrument Data Release (DR) 2 Baryon Acoustic Oscillations (BAO) data, combined with cosmic microwave background (CMB) power spectra (Planck Public Release, PR, 4) and lensing (Planck PR4 + Atacama Cosmology Telescope (ACT) Data Release (DR) 6) data, uncalibrated Type Ia Supernovae (SNe) data from Pantheon+ and Dark Energy Survey (DES) Year 5 (DESY5) samples, and Weak Lensing (WL: DES Year 1) data. The cosmological model consists of six $\Lambda$CDM parameters, and additionally, the dynamical dark energy parameters ($w_0$, $w_a$), the sum of neutrino masses ($\sum m_{\nu}$), the effective number of non-photon radiation species ($N_{\rm eff}$), the scaling of the lensing amplitude ($A_{\rm lens}$), and the running of the scalar spectral index ($\alpha_s$). Our major findings are the following:
	i) With CMB+BAO+DESY5+WL, we obtain the first 2$\sigma$+ detection of a non-zero $\sum m_{\nu} = 0.19^{+0.15}_{-0.18}$ eV (95\%). Replacing DESY5 with Pantheon+ still yields a $\sim$1.9$\sigma$ detection.
	ii) The cosmological constant lies at the edge of the 95\% contour with CMB+BAO+Pantheon+, but is excluded at 2$\sigma$+ with DESY5, leaving evidence for dynamical dark energy dataset-dependent and inconclusive.
	iii) With CMB+BAO+SNe+WL, $A_{\rm lens} = 1$ is excluded at $>2\sigma$, while it remains consistent with unity without WL data — suggesting that the existence of lensing anomaly with Planck PR4 likelihoods may depend on non-CMB datasets.
	iv) The Hubble tension persists at 3.6–4.2$\sigma$ with CMB+BAO+SNe; WL data has minimal impact.

\end{abstract}


\section{Introduction}
\label{sec:1}
The nature of dark energy (DE) remains one of the most pressing mysteries in modern cosmology. While the $\Lambda$-Cold Dark Matter ($\Lambda$CDM) model has been widely successful in explaining a range of low and high-redshift cosmological observations, the recent cosmological constraints from the Dark Energy Spectroscopic Instrument (DESI) collaboration \citep{DESI:2024mwx,DESI:2024hhd,DESI:2025zgx,DESI:2025fii} have provided tantalizing evidence for evolving dark energy, with a potential phantom crossing at a redshift of $z\simeq 0.5$ considering various popular parameterizations for the dynamical nature of dark energy \citep{DESI:2025fii}. Combined with the CMB and type Ia Supernovae observations, the DESI Data Release (DR) 2 measurements of the Baryon Acoustic Oscillations (BAO) presently rejects the cosmological constant at the level of 2.8, 3.8, and 4.2$\sigma$ depending on the supernovae dataset used (PantheonPlus \citep{Brout:2022vxf}, Union3 \citep{Rubin:2023ovl}, and Dark Energy Survey Year 5 (DESY5) Supernovae \citep{DES:2024tys} respectively), while using an 8 parameter cosmological model, with the Chevallier-Polarski-Linder (CPL) parameterization \citep{Chevallier:2000qy,Linder:2002et} for the equation of state (EoS) of the 
dynamical DE. The EoS in CPL parameterization given by $w(z)\equiv w_{0}+w_{a} ~z/(1+z)$, where $z$ is the redshift. The evidence for an evolving dark energy has strengthened slightly from the previous paper related to the first data release (DR1) of DESI \citep{DESI:2024mwx}. However, contrary to the DESI DR1 BAO results, currently even the CMB+BAO dataset combination also rejects the cosmological constant at more than 2$\sigma$ \citep{DESI:2025zgx, DESI:2025gwf}. The significant implications of these results have sparked a substantial number of subsequent studies on dark energy \citep[see, e.g.,][]{RoyChoudhury:2024wri,Tada:2024znt,Berghaus:2024kra,Park:2024jns,Yin:2024hba,Shlivko:2024llw,Cortes:2024lgw,DESI:2024kob,Carloni:2024zpl,Croker:2024jfg,Mukherjee:2024ryz,Roy:2024kni,Wang:2024dka,Gialamas:2024lyw,Orchard:2024bve,Giare:2024gpk,Dinda:2024ktd,Jiang:2024xnu,Reboucas:2024smm,Bhattacharya:2024hep,Pang:2024qyh,Ramadan:2024kmn,Wolf:2024eph,Efstathiou:2024xcq,Kessler:2025kju,Gao:2025ozb,Borghetto:2025jrk,Wolf:2025jlc,Huang:2025som,Peng:2025nez,Tsedrik:2025cwc,Reeves:2025axp,Yang:2025kgc,Chan-GyungPark:2025cri,Odintsov:2024woi,Shim:2024mrl,Gao:2024ily,Payeur:2024dnq,Ye:2024zpk,Luongo:2024zhc,Wolf:2024stt,Giare:2024ocw,Chakraborty:2025syu,Elbers:2024sha,Giare:2025pzu,Wolf:2025dss,Ye:2025ulq,Colgain:2024xqj,Colgain:2025nzf,Ye:2024ywg,Silva:2025hxw}.

Another important result from the DESI collaboration is on the neutrino masses. CMB data combined with DESI DR2 BAO puts a stringent constraint of $\sum m_{\nu} < 0.0642$ eV (95\%) in the $\Lambda$CDM+$\sum m_{\nu}$ model (assuming three degenerate neutrino masses) with no evidence for a detection of a non-zero neutrino mass sum \citep{DESI:2025zgx,DESI:2025ejh}, which also rules out the inverted mass hierarchy of neutrinos (that requires a minimum  $\sum m_{\nu}$ of 0.096 eV \citep{Esteban:2024eli}) at more than 2$\sigma$. However, given that the DESI results prefer a dynamical dark energy model over $\Lambda$CDM, it is debatable how much trust should be put on the neutrino mass bounds obtained in the  $\Lambda$CDM+$\sum m_{\nu}$ model. In a model extended with dynamical dark energy (i.e., $w_0w_a$CDM+$\sum m_{\nu}$) this bound relaxes to $\sum m_{\nu} < 0.129$ eV with CMB+BAO+DESY5, which still allowes for both the normal (that requires a minimum  $\sum m_{\nu}$ of 0.057 eV \citep{Esteban:2024eli}) and inverted hierarchies of neutrinos.\footnote{For earlier bounds on $\sum m_{\nu}$ in the literature, see e.g., \cite{RoyChoudhury:2018gay,RoyChoudhury:2019hls,Tanseri:2022zfe,Vagnozzi:2017ovm,Vagnozzi:2018jhn,RoyChoudhury:2018vnm,Sharma:2022ifr,DiValentino:2021hoh,Giusarma:2018jei,Reeves:2022aoi,Yang:2017amu}. For more recent studies, see \cite{Shao:2024mag,Jiang:2024viw, Herold:2024nvk, RoyChoudhury:2024wri,Bertolez-Martinez:2024wez,Wang:2025ker}} We note, however, that such strong bounds have attracted attention from the particle physics and cosmology community regarding the possible explanation of a lack of detection of non-zero neutrino masses and apparent possibility of the  $\sum m_{\nu}$ posterior peak occuring at a negative neutrino mass range (i.e., $\sum m_{\nu}<0$) \citep{Craig:2024tky,Lynch:2025ine,Loverde:2024nfi,Elbers:2024sha}. In $\Lambda$CDM, the DESI BAO data prefers a lower value of $\Omega_m$ than Planck (see, e.g., Fig. 10 of \cite{DESI:2025zgx}), which leads to an apparent issue of $\omega_c+\omega_b > \omega_m$ for joint analyses with CMB and BAO datasets, as pointed out in \cite{Lynch:2025ine} and \cite{Loverde:2024nfi}, and that in turn produces such strong bounds on $\sum m_{\nu}$ in the $\Lambda$CDM+$\sum m_{\nu}$ model. However, we note that the cosmological data favors an evolving dark energy instead of $\Lambda$, and in a cosmological model with evolving dark energy, the $\Omega_m$ tension does not appear \citep{DESI:2025zgx}. 

In \cite{RoyChoudhury:2024wri}, using CMB data with DESI DR 1 BAO and uncalibrated supernovae measurements, in a 12-parameter cosmological model\footnote{For previous studies in such largely extended parameter spaces, see \cite{DiValentino:2015ola,DiValentino:2016hlg,DiValentino:2017zyq,Poulin:2018zxs,RoyChoudhury:2018vnm,DiValentino:2019dzu}.}, we showed that the evidence for dynamical dark energy is not robust yet, since CMB+BAO+Pantheon+ still included the cosmological constant ($w_0 = -1$, $w_a = 0$) within 2$\sigma$ on the 2D contour plot in the $w_0-w_a$ plane. We also noticed that the $\sum m_{\nu}$ posterior probability distributions peaked in the $\sum m_{\nu}>0$ region, with three dataset combinations producing a 1$\sigma$+ detection. The extended model consisted of the six standard  $\Lambda$CDM parameters and the following simple extensions: the dynamical dark energy equation of state parameters (CPL: $w_0$ and $w_a$), the sum of neutrino masses ($\sum m_{\nu}$) and effective number of non-photon radiation species ($N_{\rm eff}$), the scaling of the lensing amplitude ($A_{\rm lens}$), and the running of the scalar spectral index ($\alpha_s$). For CMB data, we used the latest Planck Public Release 4 (PR4) likelihoods (2020): HiLLiPoP and LoLLiPoP \citep{Tristram:2023haj}, and Planck PR4 lensing combined with ACT DR6 lensing likelihoods \citep{ACT:2023kun}. For BAO, we used DESI DR1 BAO likelihoods \citep{DESI:2024mwx}, and for supernovae, the latest uncalibrated type Ia Supernovae likelihoods: Pantheon+ \citep{Brout:2022vxf} and DESY5 \citep{DES:2024tys}. 

In this paper, we extend the work in \cite{RoyChoudhury:2024wri} by using the new DESI DR2 BAO data \citep{DESI:2025zgx,DESI:2025zpo}, while using the same CMB and supernovae (SNe) datasets. Apart from that, we also use the Dark Energy Survey Year 1 (DESY1) data on galaxy clustering and weak lensing \citep{DES:2017myr}\footnote{We note here that a newer data on weak lensing from DES exists (DES Year 3), but the likelihoods are released only for CosmoSIS \citep{Zuntz:2014csq} and not Cobaya \citep{Torrado:2020dgo}, which we use in this paper.}. 

We briefly describe the reason for using the extensions to the $w_{0}w_a$CDM model as follows: Neutrinos, although massless in the Standard Model, are known to be massive from oscillation experiments, requiring at least two non-zero masses with mass-squared splittings $\Delta m_{21}^2 \simeq 7.42 \times 10^{-5}$ eV$^2$ and $|\Delta m_{31}^2| \simeq 2.51 \times 10^{-3}$ eV$^2$, leading to two possible mass orderings: Normal ($\sum m_{\nu} > 0.057$ eV) and Inverted ($\sum m_{\nu} > 0.096$ eV) \citep{Esteban:2020cvm}. Thus, the $\sum m_{\nu}$ parameter serves as a simple and natural extension to the model. Dynamical dark energy (parameterized via $w_0$, $w_a$) also impacts constraints on $\sum m_\nu$ through geometric degeneracies \citep{RoyChoudhury:2019hls}. Also note that the preferred $w_0$–$w_a$ region reported by the DESI collaboration \citep{DESI:2024mwx,DESI:2024hhd,DESI:2025zgx,DESI:2025zgx} lies in the phantom regime, which relaxes $\sum m_{\nu}$ bounds by over a factor of 2 because of the well-known degeneracy between the DE Equation of State and $\sum m_{\nu}$ \citep{Hannestad:2005gj}.

Massive neutrinos, once non-relativistic, suppress structure formation at small scales, affecting CMB lensing and enhancing small-scale anisotropy correlations \citep{RoyChoudhury:2020das, Lesgourgues:2012uu}. The $>$2$\sigma$ lensing anomaly with Planck PR3 (2018) likelihoods ($A_{\rm lens} = 1.18 \pm 0.065$ at 68\%) \citep{Planck:2018nkj, Planck:2018vyg} has nearly vanished ($<1\sigma$) in Planck PR4 \citep{Tristram:2023haj}, which justifies using Planck PR4 for robust neutrino mass inference. Since $\sum m_{\nu}$ and $A_{\rm lens}$ are correlated, including the $A_{\rm lens}$ parameter and using Planck PR4 likelihoods avoids possible systematic biases in the neutrino mass measurement.

The lensing amplitude parameter $A_{\rm lens}$ is also degenerate with the curvature density $\Omega_k$ \citep{DiValentino:2019qzk}, and the lensing anomaly is closely linked to the so-called curvature tension \citep{DiValentino:2019qzk, Handley:2019tkm}. Specifically, Planck PR3 (2018) likelihoods alone showed a preference for a closed universe, with $\Omega_k = 0$ excluded at more than 2$\sigma$. However, this curvature tension disappears once additional datasets such as BAO and supernovae are incorporated \citep{Planck:2018vyg}. DESI collaboration has also reported no significant deviation from a flat universe in a dynamical dark energy model \citep{DESI:2025zgx}. In contrast, the lensing anomaly remains significant at more than 2$\sigma$ even when these external datasets are included alongside Planck PR3. Therefore, in this work, we focus solely on the lensing anomaly and fix $\Omega_k = 0$ throughout our analysis.

The Hubble tension, a $\sim4.6\sigma$ discrepancy between SH0ES ($H_0 = 73.04 \pm 1.04$) \citep{Riess:2021jrx} and Planck PR4 ($H_0 = 67.64 \pm 0.52$) \citep{Tristram:2023haj} in $\Lambda$CDM, might be partially mitigated by phantom dark energy \citep{RoyChoudhury:2019hls, Vagnozzi:2019ezj} or a fully thermalized extra radiation species ($\Delta N_{\rm eff} \sim 1$), though the latter is disfavoured by data \citep{Planck:2018vyg, Vagnozzi:2023nrq, RoyChoudhury:2020dmd, RoyChoudhury:2022rva, Bostan:2023ped}. However, we include the $N_{\rm eff}$ parameter in our analysis to understand how far the $H_0$ discrepancy can be resolved in a 12 parameter model (even though we do not expect to fully solve the tension). A higher than standard $N_{\rm eff}$ increases the Hubble expansion rate in the pre-recombination era, which in turn increases $H_0$. Given that Hubble tension is the most pressing discrepancy in cosmology, we think the addition of $N_{\rm eff}$ as a varying parameter in the model is important. Lastly, we note that a small value of $\log_{10} |\alpha_s| = -3.2$ typically emerges in slow-roll inflationary models \citep{Garcia-Bellido:2014gna}, although some alternative inflationary scenarios can produce significantly larger values (see, e.g., \cite{Easther:2006tv,Kohri:2014jma,Chung:2003iu}). Thus, we vary $\alpha_s$ to check for any deviations from the expected value.

Our main goals for this paper are as follows: 1) As in \cite{RoyChoudhury:2024wri}, we want to check whether the evidence for dynamical dark energy survives in a largely extended parameter space with simple extensions to the cosmological model. 2) We want to re-assess the $\sum m_{\nu}$ posteriors with the new DESI DR2 BAO likelihoods and the DES Year 1 results, in this extended model, to check whether we can obtain any 2$\sigma$ detection of positive non-zero $\sum m_{\nu}$. 3) We aim to further investigate the lensing anomaly or $A_{\rm lens}$-anomaly \citep{Calabrese:2008rt,Planck:2018vyg} situation in the presence of weak lensing data. 4) We aim to assess the level of robustness of the Hubble tension \citep{Riess:2021jrx,Knox:2019rjx} in this largely extended parameter space. With the release of the DESI DR2 BAO data, we believe it is an opportune moment to revisit and update the constraints within such an extended cosmological model. The resulting constraints will undoubtedly be of significant value to both the cosmology and particle physics communities.

The structure of the paper is as follows: Section~\ref{sec:2} outlines the analysis methodology. In Section~\ref{sec:3}, we present and discuss the results of our statistical analysis. We conclude in Section~\ref{sec:4}. A summary of the cosmological parameter constraints is provided in Table~\ref{table:2}.

\section{Analysis methodology} \label{sec:2}
We outline the cosmological model, parameter sampling and plotting codes, as well as the priors on parameters in Section~\ref{sec:2.1}. Section~\ref{sec:2.2} presents a discussion on the cosmological datasets utilized in this study.

\subsection{Cosmological model and parameter sampling}\label{sec:2.1}
Here is the parameter vector for this extended model with 12 parameters :
\begin{eqnarray}\label{eqn:3}
	&\theta \equiv \left[\omega_c, ~\omega_b, ~\Theta_s^{*},~\tau, ~n_s, ~{\rm{ln}}(10^{10} A_s), \right. \nn \\ 
	& \qquad \qquad \quad w_{0, \rm DE}, w_{a, \rm DE}, 	\left.N_{\textrm{eff}}, \sum m_{\nu}, \alpha_s, A_{\textrm{lens}} \right].
\end{eqnarray}

The first six parameters correspond to the $\Lambda$CDM model: the present-day cold dark matter energy density, $\omega_c \equiv \Omega_c h^2$; the present-day baryon energy density, $\omega_b \equiv \Omega_b h^2$; the reionization optical depth, $\tau$; the scalar spectral index, $n_s$; and the amplitude of the primordial scalar power spectrum, $A_s$ (both evaluated at the pivot scale $k_* = 0.05~ \text{Mpc}^{-1}$). Additionally, $\Theta_s^{*}$ represents the ratio of the sound horizon to the angular diameter distance at the time of photon decoupling.

The remaining six parameters extend the $\Lambda$CDM cosmology. For the CPL parametrization of the dark energy equation of state, we use the notation ($w_{0,\rm DE}$, $w_{a,\rm DE}$) interchangeably with ($w_0$, $w_a$). The other parameters, as outlined in the introduction, include the effective number of non-photon radiation species ($N_{\rm eff}$), the sum of neutrino masses ($\sum m_{\nu}$), the running of the scalar spectral index ($\alpha_s$), and the scaling of the lensing amplitude ($A_{\rm lens}$).

We note that we adopt the degenerate hierarchy for neutrino masses, where all three neutrino masses are equal ($m_i = \sum m_{\nu}/3$ for $i = 1,2,3$), and impose a prior $\sum m_{\nu} \geq 0$. This choice is justified since cosmological observations primarily constrain the total neutrino mass sum through its effect on the energy density \citep{Lesgourgues:2012uu}, and even upcoming cosmological data will remain insensitive to the small neutrino mass splittings \citep{Archidiacono:2020dvx}. Furthermore, forecasts indicate that assuming a degenerate hierarchy instead of the true mass hierarchy introduces only a negligible bias in the event of a detection of $\sum m_{\nu}$ \citep{Archidiacono:2020dvx}. Additionally, there is no definitive evidence favoring a particular neutrino mass hierarchy, even when combining cosmological constraints with terrestrial experiments, such as neutrino oscillation and beta decay data \citep{Gariazzo:2022ahe}.

Since we allow for variations in the running of the scalar spectral index ($\alpha_s \equiv d n_s / d\ln k$, where $k$ is the wave number), we assume a standard running power-law model for the primordial scalar power spectrum, expressed as

\begin{equation}
\mathrm{ln}~\mathcal{P}_s (k) = \mathrm{ln}~ A_s + (n_s -1)~ \mathrm{ln}\left(\frac{k}{k_*}\right)+\frac{\alpha_s}{2} \left[ \mathrm{ln}\left(\frac{k}{k_*}\right) \right]^2.
\end{equation}

A small value of $\log_{10} |\alpha_s| = -3.2$ naturally arises in slow-roll inflationary models \citep{Garcia-Bellido:2014gna}, though certain other inflationary scenarios can yield larger values (see, e.g., \cite{Easther:2006tv,Kohri:2014jma,Chung:2003iu}).

\textbf{Parameter Sampling}: For all Markov Chain Monte Carlo (MCMC) analyses in this paper, we use the cosmological inference code Cobaya \citep{Torrado:2020dgo,2019ascl.soft10019T}. Theoretical cosmology calculations are performed using the Boltzmann solver CAMB \citep{Lewis:1999bs,Howlett:2012mh}. When incorporating the combined Planck PR4 + ACT DR6 lensing likelihood, we apply the higher precision settings recommended by ACT.

To assess chain convergence, we utilize the Gelman and Rubin statistics \citep{doi:10.1080/10618600.1998.10474787}, ensuring that all chains satisfy the convergence criterion of $R-1<0.01$. We use GetDist \citep{Lewis:2019xzd} to derive parameter constraints and generate the plots presented in this paper. Broad flat priors are imposed on the cosmological parameters, as detailed in Table~\ref{table:1}.
\begin{table}
	\begin{center}
		\begin{tabular}{c c}
			\hline
			Parameter                    & Prior\\
			\hline
			$\Omega_{\rm b} h^2$         & [0.005, 0.1]\\
			$\Omega_{\rm c} h^2$         & [0.001, 0.99]\\
			$\tau$                       & [0.01, 0.8]\\
			$n_s$                        & [0.8, 1.2]\\
			${\rm{ln}}(10^{10} A_s)$         & [1.61, 3.91]\\
			$\Theta_s^{*}$             & [0.5, 10]\\ 
			$w_{0,\rm DE}$                        & [-3, 1]\\
			$w_{a,\rm DE}$                        & [-3, 2]\\
			$N_{\rm eff}$                & [2, 5]\\ 
			$\sum m_\nu$ (eV)            & [0, 5]\\
			$\alpha_s$                    & [-0.1, 0.1]\\
			$A_{\textrm{lens}}$          & [0.1, 2]\\
			\hline
		\end{tabular}
	\end{center}
	\caption{\label{table:1} Flat priors on the main cosmological parameters constrained in this paper.}
\end{table}

\subsection{Datasets} \label{sec:2.2}
\textbf{CMB: Planck Public Release (PR) 4}: We utilize the most recent large-scale (low-$l$) and small-scale (high-$l$) Cosmic Microwave Background (CMB) temperature and E-mode polarization power spectra measurements from the Planck satellite. For the high-$l$ ($30<l<2500$) TT, TE, and EE data, we adopt the latest HiLLiPoP likelihoods, as detailed in \cite{Tristram:2023haj}. The low-$l$ ($l<30$) EE spectra are analyzed using the most recent LoLLiPoP likelihoods, also described in \cite{Tristram:2023haj}. Both of these likelihoods are derived from the Planck Public Release (PR) 4, the latest reprocessing of data from the LFI and HFI instruments through a unified pipeline, NPIPE, which provides a slightly larger dataset, reduced noise, and improved consistency across frequency channels \citep{Planck:2020olo}. For the low-$l$ TT spectra, we employ the Commander likelihood from the Planck 2018 collaboration \citep{Planck:2018vyg}. We collectively refer to this set of likelihoods as \textbf{``Planck PR4."}

\textbf{CMB lensing: Planck PR4+ACT DR6}. CMB experiments also measure the power spectrum of the gravitational lensing potential, $C_l^{\phi \phi}$, through 4-point correlation functions. In our analysis, we utilize the latest NPIPE PR4 Planck CMB lensing reconstruction \citep{Carron:2022eyg} along with the Data Release 6 (DR6) from the Atacama Cosmology Telescope (ACT) (version 1.2) \citep{ACT:2023kun,ACT:2023ubw}. Following the recommendations of the ACT collaboration, we adopt the higher precision settings \citep{ACT:2023kun}. For conciseness, we refer to this dataset combination as \textbf{``lensing"}. 

\textbf{BAO: DESI Data Release (DR) 2}.We incorporate the latest measurement of the Baryon Acoustic Oscillation (BAO) signal from Data Release 2 of the Dark Energy Spectroscopic Instrument (DESI) collaboration \citep{DESI:2025zgx,DESI:2025zpo} (for reference to the earlier DR1, see \cite{DESI:2024mwx}). This dataset includes observations from the Bright Galaxy Sample (BGS, $0.1 < z < 0.4$), the Luminous Red Galaxy Sample (LRG, $0.4 < z < 0.6$ and $0.6 < z < 0.8$), the Emission Line Galaxy Sample (ELG, $1.1 < z < 1.6$), the combined LRG and ELG sample within a shared redshift range (LRG+ELG, $0.8 < z < 1.1$), the Quasar Sample (QSO, $0.8 < z < 2.1$), and the Lyman-$\alpha$ Forest Sample (Ly$\alpha$, $1.77 < z < 4.16$). We refer to this complete dataset as \textbf{``DESI2."}

\textbf{SNe~Ia: Pantheon+}.We incorporate the latest Supernovae Type-Ia (SNeIa) luminosity distance measurements from the Pantheon+ Sample \cite{Scolnic:2021amr}, which consists of 1550 spectroscopically confirmed SNeIa spanning the redshift range $0.001 < z < 2.26$. For our analysis, we use the publicly available likelihood from \cite{Brout:2022vxf}, which accounts for both statistical and systematic covariance. This likelihood applies a constraint of $z > 0.01$ to mitigate the impact of peculiar velocities on the Hubble diagram. We refer to this dataset as \textbf{``PAN+"}.

\textbf{SNe~Ia: DES Year 5}. We make use of the luminosity distance measurements from the latest supernova sample, which includes 1635 photometrically classified SNeIa in the redshift range $0.1<z<1.13$, publicly released by the Dark Energy Survey (DES) as part of their Year 5 data release \citep{DES:2024tys}. We refer to this dataset as \textbf{``DESY5"}.

We note that PAN+ and DESY5 share some supernovae in common. To prevent double counting, these two datasets are never used simultaneously in our analysis.

\textbf{Weak Lensing: DES Year 1}. We include the likelihood from the combined analysis of galaxy clustering and weak gravitational lensing, using 1321 deg$^2$  of \textit{griz} imaging data from the first year of the Dark Energy Survey \citep{DES:2017myr}. We refer to this dataset as \textbf{``WL"}.

\section{Numerical results} \label{sec:3}
The main findings from our cosmological parameter estimation are summarized in Table~\ref{table:2} and illustrated in Figures~\ref{fig:1}–\ref{fig:6}.

\begin{table*}
	
	\centering
	\resizebox{\textwidth}{!}{
		\begin{tabular}{ccccccc}
			\toprule
			\toprule
			\vspace{0.2cm}
			Parameter  &    Planck PR4  &    Planck PR4  &    Planck PR4  &    Planck PR4  &    Planck PR4   \\\vspace{0.2cm}
			&   +lensing+DESI2   & +lensing+DESI2+PAN+ &  +lensing+DESI2+PAN++WL &  +lensing+DESI2+DESY5 &    +lensing+DESI2+DESY5+WL     \\
			\midrule
			\hspace{1mm}
			\vspace{ 0.2cm}
			$\Omega_b h^2$  &    $0.02238\pm0.00020$  &    $0.02246\pm0.00019$  &    $0.02253\pm0.00019$  &    $0.02242\pm0.00020$  &    $0.02251\pm0.00019$    \\ \vspace{ 0.2cm}
			
			$\Omega_c h^2$  &    $0.1190\pm0.0028$  &    $0.1191\pm0.0028$  &    $0.1184\pm0.0028$  &    $0.1190\pm0.0028$  &    $0.1184\pm0.0028$    \\ \vspace{ 0.2cm}
			
			$\tau$  &    $0.0584^{+0.0061}_{-0.0068}$  &     $0.0586\pm0.0066$  &    $0.0579\pm0.0065$  &    $0.0586\pm0.0065$  &      $0.0579 \pm 0.0065$    \\ \vspace{ 0.2cm}
			
			$n_s$  &    $0.972\pm0.009$  &    $0.975\pm0.009$  &    $0.978\pm0.009$  &     $0.974\pm0.009$ &   $0.977\pm0.009$  \\ \vspace{ 0.2cm}
			
			${\rm{ln}}(10^{10} A_s)$  &    $3.042\pm0.016$  &    $3.043\pm0.016$  &    $3.039\pm0.016$  &   $3.043\pm0.016$   &     $3.039\pm0.016$    \\ \vspace{ 0.2cm}
			
		100$\Theta_s^{*}$  &    $1.04082\pm0.00040$  &     $1.04082\pm0.00039$  &    $1.04086\pm0.00040$   &     $1.04083\pm0.00039$  &  $1.04085\pm0.00039$ \\ 

		\multirow{2}{*}{$\sum m_\nu$ (eV)}  &    $0.147^{+0.064}_{-0.12}$ (1$\sigma$),  &    \multirow{2}{*}{$<0.242$ (2$\sigma$)}  &    $0.166\pm0.087$ (1$\sigma$),  &     \multirow{2}{*}{$<0.261$ (2$\sigma$)}  &    $0.190\pm0.088$ (1$\sigma$),   \\ \vspace{ 0.2cm}
		& $<0.302$ (2$\sigma$)&& $<0.313$ (2$\sigma$) &&   $0.19^{+0.15}_{-0.18}$ (2$\sigma$)
		\\ \vspace{ 0.2cm}
			
			$N_{\textrm{eff}}$  &    $3.10\pm0.19$  &    $3.15\pm 0.19$  &    $3.16\pm0.19$  &    $3.12\pm0.19$  &    $3.15\pm0.19$   \\ \vspace{ 0.2cm}
			
			$w_0$  &    $-0.46\pm0.23$  &    $-0.864\pm0.056$  &    $-0.859\pm0.057$  &    $-0.775\pm0.061$ &   $-0.768\pm0.062$  \\ \vspace{ 0.2cm}
			
			$w_a$  &    $-1.61\pm0.70$  &    $-0.44^{+0.26}_{-0.22}$  &    $-0.47^{+0.27}_{-0.23}$  &    $-0.72^{+0.29}_{-0.24}$  &    $-0.76^{+0.30}_{-0.26}$   \\ \vspace{ 0.2cm}

			$n_{\rm run}$  &    $-0.0031\pm0.0074$  &     $-0.0018\pm0.0072$  &     $0.0004\pm0.0072$ &   $-0.0022\pm0.0073$  &    $0.0002\pm0.0072$  \\ 
			
				\multirow{2}{*}{$A_{\textrm{lens}}$}  &    \multirow{2}{*}{$1.061^{+0.046}_{-0.054}$ (1$\sigma$)} &    \multirow{2}{*}{$1.068^{+0.042}_{-0.050}$ (1$\sigma$)}    &    $1.104\pm0.044$ (1$\sigma$),   &    \multirow{2}{*}{$1.063^{+0.043}_{-0.052}$ (1$\sigma$)}   &    $1.104\pm0.044$ (1$\sigma$),    \\ \vspace{ 0.2cm} 
				& &  &  $1.104^{+0.089}_{-0.084}$ (2$\sigma$) &  & $1.104^{+0.090}_{-0.085}$ (2$\sigma$) 
				\\ 
			\midrule
		
		   	$H_0$ (km/s/Mpc)  &    $64.0^{+2.0}_{-2.6}$  &    $67.9\pm1.0$  &    $67.9\pm1.0$   &     $67.0\pm1.0$   &    $67.1\pm1.0$   \\ \vspace{ 0.2cm}
		   
		   $S_8$  &    $0.823\pm0.021$  &    $0.808^{+0.019}_{-0.016}$  &    $0.791\pm0.015$  &    $0.812^{+0.019}_{-0.017}$  &    $0.793\pm0.016$   \\ \vspace{ 0.2cm}
		   $\Omega_m$  &    $0.350\pm0.023$  &    $0.309\pm0.006$  &     $0.309\pm0.006$   &     $0.318\pm0.006$   &    $0.318\pm0.006$    \\

			\bottomrule
			\bottomrule

	\end{tabular}}
	\caption{\label{table:2}\footnotesize Bounds on cosmological parameters in the 12 parameter extended model. Marginalized limits are given at 68\% C.L. whereas upper limits are given at 95\% C.L. Note that $H_0$, $S_8$, and $\Omega_m$ are derived parameters.}
\end{table*}

\subsection{$w_{0,\rm DE}$ and  $w_{a,\rm DE}$}
\begin{figure}[tbp]
	\centering 
	\includegraphics[width=.95\linewidth]{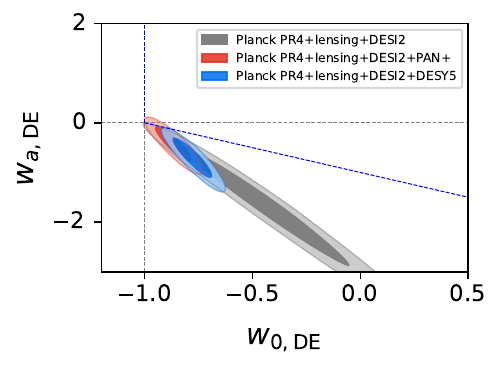}
	\caption{\label{fig:1} 68\% and 95\% marginalised  contours in the $w_{0,\rm DE} - w_{a,\rm DE}$ plane for different data combinations. The area to the right of the vertical dashed blue line and above the slanted dashed blue line represents the parameter space corresponding to quintessence-like or non-phantom dark energy.}
\end{figure}

\begin{figure*}[tbp]
	\centering 
	\includegraphics[width=.48\linewidth]{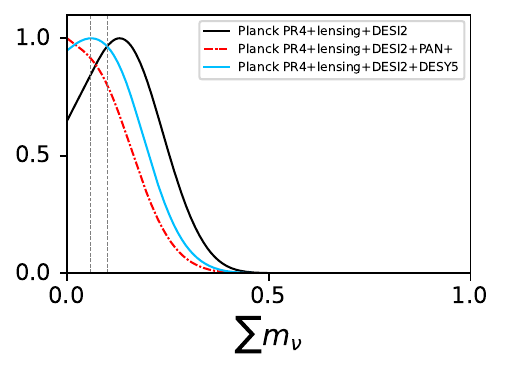}
	\hfill
	\includegraphics[width=.48\linewidth]{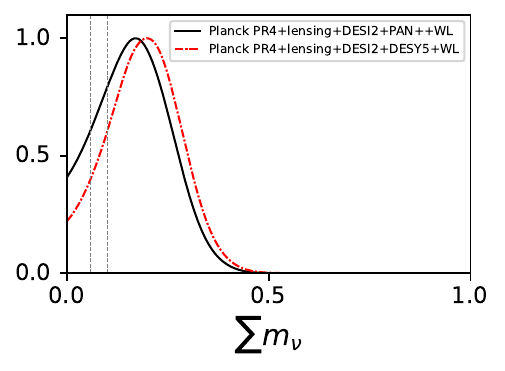}
	\caption{\label{fig:2}A comparison of the 1D marginalized posterior distributions for $\sum m_{\nu}$ [eV] across various data combinations. The panel in the right shows results with the DES Year 1 Weak Lensing data (WL) included. Note that in the right panel, the Planck PR4+lensing+DESI2+DESY5+WL dataset combination leads to a 2$\sigma$+ detection of non-zero $\sum m_{\nu}$. The two vertical black dashed lines (in both panels) indicate the minimum mass thresholds for the normal (0.057 eV) and inverted (0.096 eV) neutrino mass hierarchies, respectively.}
\end{figure*}

\begin{figure*}[tbp]
	\centering 
	\includegraphics[width=.32\linewidth]{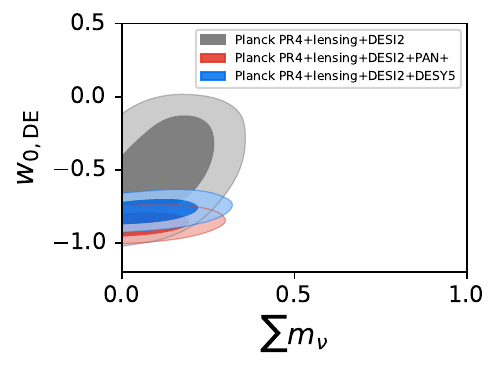}
	\hfill
	\includegraphics[width=.32\linewidth]{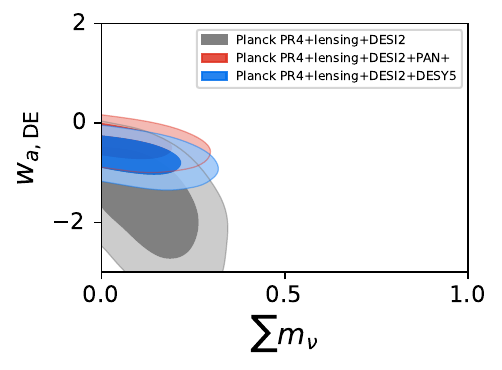}
	\hfill
	\includegraphics[width=.32\linewidth]{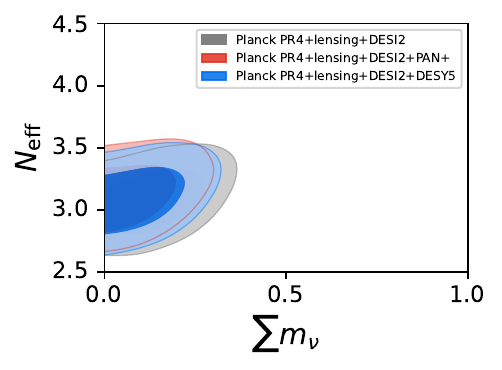}
	\vfill
	\includegraphics[width=.32\linewidth]{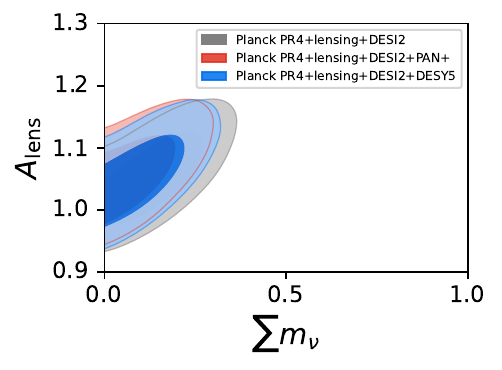}
	\hfill
	\includegraphics[width=.32\linewidth]{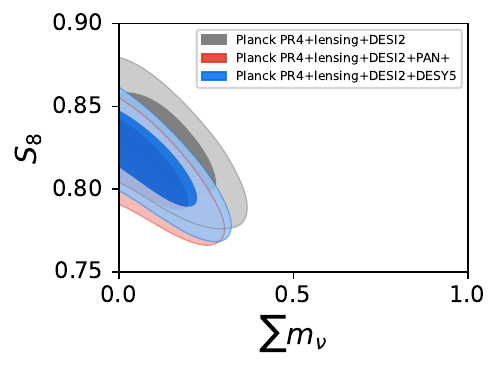}
	\hfill
	\includegraphics[width=.32\linewidth]{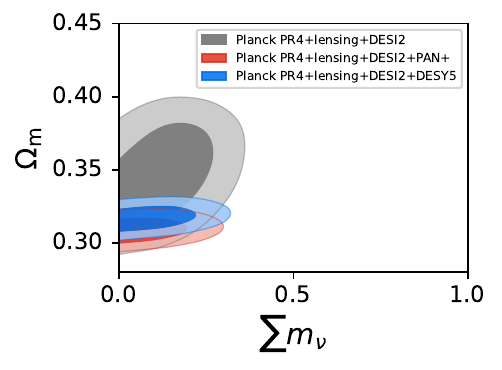}
	\caption{\label{fig:3}A comparison of the 2D correlation plots between $\sum m_{\nu}$ [eV] and various other parameters across various data combinations.}
\end{figure*}

	 As shown in Figure~\ref{fig:1}, when CMB and BAO data are combined with Pantheon+, the cosmological constant scenario ($w_{0,\rm DE} = -1$, $w_{a,\rm DE} = 0$) lies at the edge of the 95\% confidence contour. We also find that a region of the quintessence/non-phantom dark energy parameter space ($w(z)\geq -1$ at all redshifts) is also allowed within 2$\sigma$ with Pantheon+. However, when using the DESY5 SNe~Ia data, we find that the cosmological constant is excluded at more than 2$\sigma$, with a $\sim$2$\sigma$ level tension also observed for non-phantom (quintessence-like) dark energy models.  Therefore, we conclude that the evidence reported by the DESI BAO collaboration for a dynamical dark energy equation of state is not yet conclusive.  We note that the addition of the WL data has negligible effect on the constraints on the DE equation of state parameters, and thus it does not change any conclusions regarding the same.
	 
	  We also note that recent studies have reported a correlation between $A_{\rm lens}$ and the DE Equation of State parameters \citep{Chan-GyungPark:2025cri,Park:2024pew}, which relaxes the bounds on the $w_0$, $w_a$ parameters. We had seen this relaxation in an earlier paper as well \citep{RoyChoudhury:2024wri}. This correlation in turn, reduces the tension with $\Lambda$CDM compared to what is reported by the DESI collaboration paper \citep{DESI:2025zgx}. On the other hand, the reduction in tension with $\Lambda$CDM is propagating to an increase in the value of $A_{\rm lens}$, and thus, the $A_{\rm lens}$ parameter no longer remains in agreement with $A_{\rm lens} = 1$ at less than 1$\sigma$ (even though it remains in agreement within 2$\sigma$ with CMB+BAO+SNe).

\subsection{$\sum m_{\nu}$} 
Figure~\ref{fig:2} shows the 1D marginalized posterior distributions of $\sum m_{\nu}$ for different dataset combinations, while Figure~\ref{fig:3} displays the 68\% and 95\% 2D confidence contours between $\sum m_{\nu}$ and several other cosmological parameters. We notice in the left panel of Figure~\ref{fig:2}, that without the WL data, there is no 2$\sigma$ detection of non-zero neutrino masses, but still, we do find that the posteriors peak in the $\sum m_{\nu} > 0$ region for Planck PR4+lensing+DESI2 and Planck PR4+lensing+DESI2+DESY5. In fact, there is a 1$\sigma$+ detection with Planck PR4+lensing+DESI2. However, with the addition of the WL data, we find clear peaks in the $\sum m_{\nu}$ posteriors, with a 2.1$\sigma$ detection with Planck PR4+lensing+DESI2+DESY5+WL at $\sum m_{\nu} = 0.19^{+0.15}_{-0.18}$ eV and a 1.9$\sigma$ detection with Planck PR4+lensing+DESI2+PAN++WL. As far as we are aware, this is the first 2$\sigma$+ detection of a non-zero $\sum m_{\nu}$ with DESI DR2 BAO data. Also, we note that the $\sum m_{\nu} > 0$ prior used in this paper might lead to slightly larger $\sum m_{\nu}$ than a model incorporating a prior free scenario with an effective neutrino mass sum that allows the negative masses \citep{DESI:2025ejh}. 

The detection with WL follows from the strong negative correlation between $S_8$ and $\sum m_{\nu}$, as visualized in the middle bottom panel of Figure~\ref{fig:3}. The observed shift of the posteriors in this paper towards more positive values is driven by the small $S_8$-tension present between weak lensing (WL) and the CMB+BAO+SNe. Since WL favors a lower clustering amplitude, this discrepancy can be addressed by either decreasing $\sigma_8$, increasing the neutrino mass $m_\nu$, or alternatively modifying the matter density parameter $\Omega_m$ \citep{DES:2017myr}. The lower $S_8$ values due to the WL data leads to rejection of smaller neutrino masses. We note here that the DES Year 1 WL data used in this work provides a value of $S_8 = 0.783^{+0.021}_{-0.025}$ in the $\Lambda$CDM model \cite{DES:2017myr}, which is only discrepant at the level of 0.95$\sigma$ with the $S_8$ value in this 12 parameter model using CMB+BAO+PAN+ and 1.07$\sigma$ with CMB+BAO+DESY5. Thus it is okay to combine the WL data with the CMB+BAO+SNe combination. Also note that if we consider a 12 parameter model with WL data alone, then the errors on $S_8$ will likely be larger, thereby reducing the discrepancy further. Previously in \cite{RoyChoudhury:2024wri}, we had also noted that the $S_8$ tension is only at the level of 1.4$\sigma$ with the DES Year 3 data (see also \cite{Tristram:2023haj}), which prefers slightly lower values of $S_8$ than DES Year 1 \citep{DES:2021wwk}. Therefore, had we used the DES Year-3 likelihoods instead, we would likely have obtained stronger evidence for a non-zero $\sum m_{\nu}$. It is worth noting, however, that recent results from the completed KiDS survey report slightly higher constraints on $S_8$, with $S_8 = 0.815^{+0.016}_{-0.021}$ within the $\Lambda$CDM framework \citep{Wright:2025xka}. Therefore, it remains uncertain whether a non-zero $\sum m_{\nu}$ detection would persist when using the KiDS dataset. However, caution is warranted, as the model favored by the DESY5 data deviates significantly from the standard $\Lambda$CDM framework, and thus, the $S_8$ inference from KiDS data might also differ from $\Lambda$CDM. We also observe that with the CMB+BAO+SNe combination, $\sum m_{\nu}$ exhibits mild correlations with $N_{\rm eff}$, $\Omega_m$, and $w_0$, while its correlations with $A_{\rm lens}$, $w_a$, and $S_8$ are notably stronger. These parameter correlations contribute to the loosening of the constraints on $\sum m_{\nu}$.

\begin{figure}[tbp]
	\centering 
	\includegraphics[width=.95\linewidth]{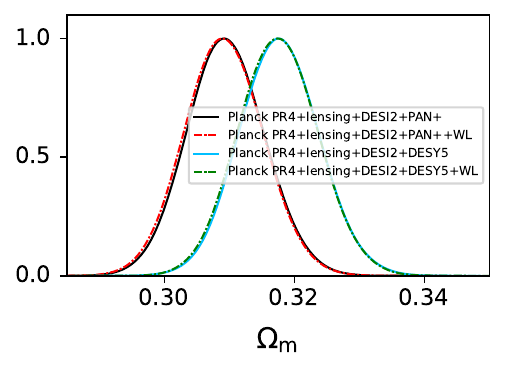}
	\hfill
	\includegraphics[width=.95\linewidth]{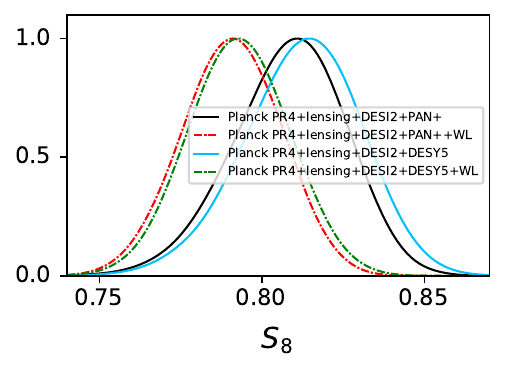}
	\caption{\label{fig:4}A comparison of the 1D marginalized posterior distributions for $\Omega_m$ and $S_8$ across various data combinations. Note that the $\Omega_m$ posteriors remain similar with the addition of WL data, but the $S_8$ values are lowered.}
\end{figure}

\begin{figure}[tbp]
	\centering 
	\includegraphics[width=.95\linewidth]{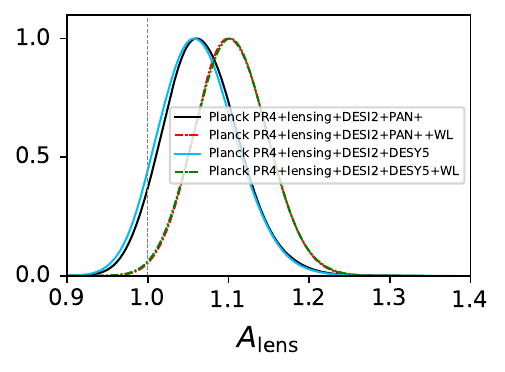}
	\hfill
	\includegraphics[width=.95\linewidth]{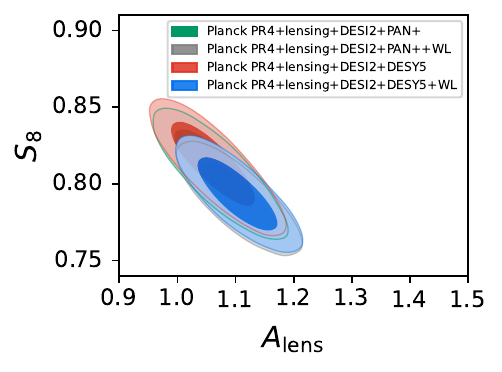}
	\caption{\label{fig:5} The left panel shows the 1D posterior distributions of $A_{\rm lens}$ for various data combinations. The right panel shows its 2D correlation plots with the $S_8$ parameter. We note that dataset combinations with WL included leads to a 2$\sigma$+ lensing anomaly due to the strong correlation with $S_8$.}
\end{figure}

\begin{figure*}[tbp]
	\centering 
	\includegraphics[width=.3\linewidth]{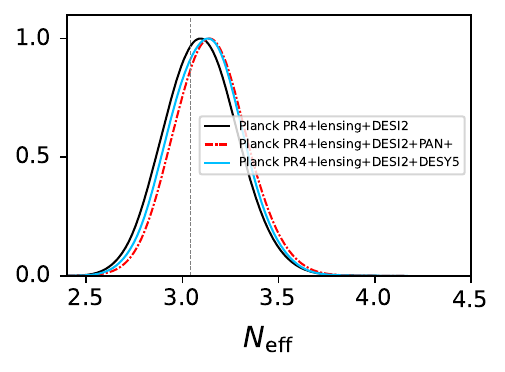}
	\hfill
	\includegraphics[width=.3\linewidth]{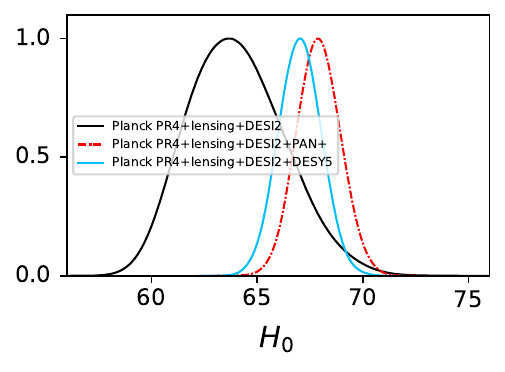}
	\hfill
	\includegraphics[width=.3\linewidth]{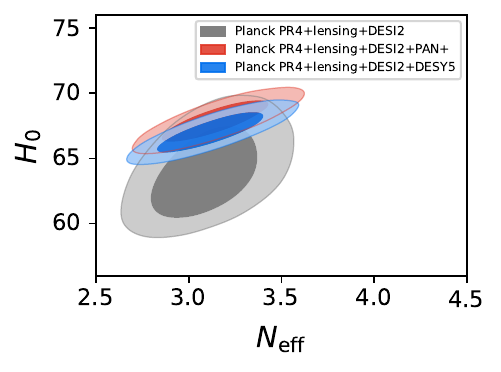}
	\caption{\label{fig:6} The left and middle panels show the 1D posterior distributions of $N_{\rm eff}$ and $H_0$ (km/s/Mpc) respectively, for various data combinations. The dashed vertical line in the left panel corresponds to the standard model value of $N_{\rm eff} = 3.044$. The right panel shows the 2D correlation plot between the two parameters, showing a strong correlation between them.}
\end{figure*}

\subsection{$\Omega_m$ and $S_8$}
 From the 1D posterior distributions in the left panel of Figure~\ref{fig:4}, we find that while the inclusion of WL data has minimal effect on the estimation of $\Omega_m$, the DESY5 dataset favors slightly higher values of $\Omega_m$ compared to Pantheon+. In contrast, the right panel of Figure~\ref{fig:4} shows that, as expected, adding the WL data shifts the preferred $S_8$ values to lower values.

\subsection{$A_{\rm lens}$} 
The right panel of Figure~\ref{fig:5} reveals a strong negative correlation between $S_8$ and $A_{\rm lens}$. Examining the left panel, we observe that the inclusion of WL data shifts the $A_{\rm lens}$ values to higher values—a direct consequence of this correlation. As shown in Table~\ref{table:2}, with the addition of WL data, the inferred $A_{\rm lens}$ deviates from unity by more than 2$\sigma$. This is a significant result regarding the Planck PR4 likelihoods as we find proof that the existence of the lensing anomaly at greater than 2$\sigma$ levels with Planck PR4 likelihoods might depend on non-CMB datasets as well, like the WL data here. Previous work with older Planck likelihoods had shown the possible exacerbation of the already present >2$\sigma$ lensing anomaly with weak lensing data \citep{Renzi:2017cbg}.

\subsection{$N_{\rm eff}$ and $H_0$}
 From the left-most panel of figure~\ref{fig:6}, we find that the obtained posteriors of $N_{\rm eff}$ are in complete agreement with the standard model value of $N_{\rm eff} = 3.044$. Whereas, from the middle panel, we notice that the $H_0$ values are not high enough to solve the Hubble tension. Indeed, if one uses the values of $H_0$ from table~\ref{table:2}, one finds that the Hubble tension is present at the level of 3.6-4.2$\sigma$ depending on the supernovae dataset used. Thus, one can consider that the Hubble tension is robust against the simple extensions to $\Lambda$CDM studied in this paper. Thus, the simple extensions to $\Lambda$CDM studied in this paper to deal with the $H_0$-tension are not sufficient to solve the Hubble tension below 2$\sigma$ level. Addition of the WL data does not change these numbers significantly. The right-most panel in Figure~\ref{fig:6} shows the expected strong correlation between $N_{\rm eff}$ and $H_0$.

\section{Conclusions} \label{sec:4}
Building upon our previous work in \cite{RoyChoudhury:2024wri}, in this paper, we have presented updated cosmological constraints within a 12-parameter extended cosmological model, utilizing a comprehensive combination of recent datasets. These include Baryon Acoustic Oscillations (BAO) from the DESI Data Release 2, Cosmic Microwave Background (CMB) temperature and polarization power spectra from Planck PR4, and CMB lensing data from Planck PR4+ACT DR6, uncalibrated type Ia Supernovae (SNe) from both the Pantheon+ and DES Year 5 (DESY5) surveys, and Weak Lensing (WL) measurements from the DES Year 1 survey. The parameter space extends the standard six $\Lambda$CDM parameters by including the dark energy equation of state parameters ($w_0$, $w_a$), the sum of neutrino masses ($\sum m_{\nu}$), the effective number of non-photon relativistic species ($N_{\rm eff}$), the lensing amplitude scaling ($A_{\rm lens}$), and the running of the scalar spectral index ($\alpha_s$). Our key results are summarized as follows:
\begin{itemize}
	\item \textbf{Neutrino Mass Detection:} Using CMB+BAO+DESY5+WL, we report the first 2$\sigma$+ preference for non-zero neutrino mass with $\sum m_{\nu} = 0.19^{+0.15}_{-0.18}$ eV (95\%). A similar, though slightly weaker, $\sim$1.9$\sigma$ signal is obtained when DESY5 is replaced with Pantheon+. Without the WL dataset, while there is no significant detection of non-zero neutrino masses, we still find that the $\sum m_{\nu}$ posteriors peak at the $\sum m_{\nu}>0$ region for CMB+BAO and CMB+BAO+DESY5. The detection with WL data is driven by a strong negative correlation between $S_8$ and $\sum m_{\nu}$, with lower $S_8$ values preferring larger masses. We note that there is no significant $S_8$-tension between WL and CMB+BAO+SNe, thus it is okay to combine them. 

\item \textbf{Dynamical Dark Energy:} We find that the cosmological constant lies at the edge of the 95\% confidence contour when using CMB+BAO+Pantheon+, and is excluded at more than 2$\sigma$ when DESY5 is included instead of Pantheon+. This suggests that the evidence for dynamical dark energy still remains dataset-dependent and inconclusive. We note that a region of the quintessence/non-phantom dark energy is also allowed by the datasets when we use Pantheon+. Addition of the WL data has negligible impact on the dynamical dark energy constraints. 

\item \textbf{Lensing Anomaly:} We find that $A_{\rm lens} = 1$ is excluded at over 2$\sigma$ when WL data is included alongside CMB+BAO+SNe. In contrast, without WL, the results remain consistent with $A_{\rm lens} = 1$ at 2$\sigma$ (albeit not at 1$\sigma$). This indicates that in regards to the Planck PR4 likelihoods, the existence of lensing anomaly at greater than 2$sigma$ level might be dependent on non-CMB datasets, such as galaxy weak lensing measurements.

\item \textbf{Hubble Tension:} The Hubble tension remains unresolved, with a persistent 3.6–4.2$\sigma$ discrepancy between CMB+BAO+SNe and the SH0ES measurement \citep{Riess:2021jrx}, depending on the SNe dataset used. We conclude that the simple extensions to $\Lambda$CDM studied in this paper are not enough to effectively solve the Hubble tension (i.e., below 2$\sigma$ level). Thus, one may need other extensions not addressed in this paper, such as modifications to pre-recombination physics and/or post-recombination late-time physics like dark energy-dark matter coupling \citep{DiValentino:2021izs,Vagnozzi:2023nrq}. The addition of WL data does not significantly alter this tension. 

\end{itemize}
Overall, our analysis emphasizes the critical importance of combining multiple cosmological probes and of testing large extensions to the standard model of cosmology to obtain a better understanding of cosmological parameters. While hints of physics beyond $\Lambda$CDM continue to appear in individual sectors—such as neutrino masses, lensing amplitude, and dark energy—their statistical significance remains sensitive to dataset combinations.

\acknowledgments
We acknowledge the use of the HPC facility at ASIAA where the numerical analyses were done. S.R.C. also acknowledges the support from grant Nos. AS-IA-112-M04, NSTC 112-2112-M-001-027-MY3, and I-IAA-ROY.

\bibliographystyle{apj}
\bibliography{biblio}




\end{document}